25 October 1997

# TOWARDS SOLUTION OF THE GAMMA-RAY BURST MYSTERY
## II. "Failed- Type I supernova/hypernova" in high-redshift starburst galaxies

by
Anup Rej & Erlend Østgaard

Anup.Rej @ phys.ntnu.no
Erlend.Oestgaard@phys.ntnu.no

Institutt for Fysikk, Lade, NTNU, 7034 Trondheim

**ABSTRACT**

We study the scenario where the gamma-ray bursts (GRBs) arise from core-collapse of very massive stars in star-forming regions in the starburst galaxies at high redshift. The bimodial structure of the gamma-ray bursts point to their association with the two redshift peaks observed in the recent study of high-redshift galaxies; one of these regions involves massive star-forming galaxies of normal type at z≈3, the other is related to the star formation in dwarf galaxies around z≈1. The cosmological time-dilation observed in GRBs is consistent with this picture and the time-dilation observed in high-redshift Type I supernovae. The details of the GRB characteristics, the nature of the starburst galaxies, and a "failed-TypeI supernova/hypernova" model of GRBs are discussed in several papers. In this paper II we discuss the details of star-formation in starburst galaxies both in the local universe and at high redshift, and we study the association of these star-forming regions with GRBs. The details of the "failed-supernova/hypernova" model of GRBs will be discussed in the next article (paper III). According to this model, the collapse of a massive star produces a dense torus of circumstellar material as well as a relativistic shock-wave accompanied by ejecta, which generate shells by impact with the material around the collapsed core, which in turn cause high-energy gamma-ray emission and the observed afterglow.

# I. INTRODUCTION

Star bursts are brief episodes of star formation that occur in the central-most $10^2$-$10^3$ pc-scale regions of starburst galaxies, and dominate the overall luminosity of such galaxies. In these galaxies star formation rates are so high that the existing gas supply can sustain starbursts for only a small fraction of the age of the universe.[1]

Starbursts are believed to be powered by high-mass stars (>8 solar masses) and are very luminous (more than a few thousand times that of the sun). They are also short-lived ( from a few million years to a few tens of millions of years) and trace relatively recent star-formation.

In the local universe, high-mass stars are found both in normal galaxies like our own Milky Way, and in local starburst galaxies. In normal galaxies these stars are distributed in the spiral arms throughout a ~ 30 kpc scale disk. In such disks the existing supply of gas can sustain the current rate of star-formation for many Gyrs. In starburst galaxies the high-mass stars are concentrated in a small region in the galactic centre (~ 100 times smaller than the galaxy as a whole), and create transient events of star-formation with a duration $< 10^8$ years.

In the local universe, more than half of the high-mass star formation comes from the nuclear region of the galaxies. Within 10 Mpc distance from our Galaxy, 25% of the high-mass star formation is accounted for by less than a handful of starburst galaxies. At intermediate redshifts, the bulk of the excess faint blue galaxy counts could be due to a dwarf galaxy population, which is experiencing high-efficiency fast evolving starbursts. At higher redshifts the galaxies indicate a high intensity star-forming phase.

These high-mass stars are very hot (>25 000 K), and a luminous output of a starburst galaxy has its peak in the vacuum-ultraviolet range (~ 912 - 3000 Å). Starbursts are also rich in interstellar gas and dust. The dust grains absorb much of the ultraviolet and are heated to temperatures of order 10-100 K, before they cool by emitting far-infrared radiation.

The high-mass stars are also dominant sources of photons, which are sufficiently energetic to photo-ionize hydrogen in the interstellar medium. When this ionized hydrogen recombines and the captured electrons cascade to the ground state, it emits photons. The luminosity of the hydrogen emission lines provides a measure of the total number of high-mass stars in the Galaxy.

The sample of increasingly luminous starburst galaxies gives morphological evidence that they are interacting or merging systems. They show presence of two or more nuclei within a single distorted envelope, or presence of a second galaxy with requisite proximity and relative brightness, together with bridges and long linear tails that are hallmarks of tidally interacting disks. The most powerful starbursts are associated with the evolution of elliptical galaxies via mergers.

## II. STARBURST GALAXIES AT VERY HIGH REDSHIFTS

In the last few years a number of deep redshift surveys[2] and deep HST imaging have studied the evolutionary state of galaxies at z ≤ 1.5. The later and less bright types of luminosity functions have undergone a significant amount of evolution in the number density and/or luminosity (of star-formation rate) over the cosmic epochs, while during the same period the bright and more massive galaxies have remained substantially unchanged.

Similar to bright galaxies at the present epochs, the high-redshift galaxies are characterised by strong clustering . HST imaging of z>3 shows that these objects are bright galaxies in an epoch of relatively intense star formation. Most of the z>3 galaxies are characterized by compact morphology, generally having a core ≤ 1.5 arcsec in diameter. The core typically contains about 90-95% of the total luminosity and has a half-light radius in the range 0.2-0.3 arcsec.

The high-redshift galaxies appear to be scaled-up (larger and luminous) versions of the local starbursts. Moreover, the average surface mass-density of stars within a half-light radius (~ $10^2$-$10^3$ $M_\odot$ pc$^{-2}$) is quite similar to the values in present day ellipticals.

The giant ellipticals and spirals at z≈ 1 are already mature slowly-evolving galaxies, with perhaps larger star-formation than at z ~ 0. Broadly speaking, massive galaxies at z ~ 1 are similar to galaxies at z ~ 0. Only the sub-L* population undergoes rapid evolution at z<1. There is a steady increase in the mass of actively star-forming systems with an increasing redshift[3]: Out to z ≈ 0.4 only 0.01-0.1 L* systems undergo significant star formation, while in the range 0.8 < z <1.6 the systems that are more massive ( in some case approaching L*) are observed to be in formation. The local starburst systems have generally low masses. The rapidly evolving population of faint blue galaxies which dominates the less massive systems at moderate redshifts, undergoes bursts of star formation at enhanced rates compared to the local starbursts.

Extremely red objects with R > 26 mag (colours of (R-K) > $5^m$), once believed to be luminous star-forming galaxies at redshifts z >6, have been found to be ellipticals at z ≈ 0.8. More EROs have been found around high-z QSOs, although they largely appear to be foreground objects. EROs are mostly compact, but sometimes resolved. Spectroscopy of HR 10 [4] indicates an evolved dusty system at modest redshift z =1.44. The z>3 galaxies from the 0000-263, 0347-383 and SSA22(2217-003) fields show that they are mostly compact galaxies, typically characterized by a bright core less than 1.5 arcsec in diameter, often surrounded by a more diffuse nebulosity with significantly lower surface brightness. The nebulosity extends to larger areas and is more irregularly distributed. Since the emission is directly proportional to the formation rate of the massive stars at the observed rest-frame in far-UV wavelengths, 90-95% of the stars which are being formed in these galaxies are concentrated in a region whose size is that of a present-day luminous galaxy. Moreover, the characteristics of the central concentration of the star formation has a size similar to a present-day spheroid. If the morphology of the massive stars is a good tracer of the overall stellar distribution, then the light distribution of the core is consistent with a dynamically relaxed structure. The central SB of the compact z>3 galaxies is consistently close to 23 mag arcsec$^{-2}$. In a few cases hierarchical merging systems are found. Sub-units merge into more massive systems, that take place in time scales about an order of magnitude shorter than the time-stretch of the probed cosmic epoch. In all cases the "merging" units have smaller luminosity than the other systems, but comparable surface brightness. Both parents and daughters of this possible merging scenario have comparable morphologies.

The photometric redshifts for the Hubble Deep Field survey show two peaks[5], one at z ~ 0.5 and another at z ~ 2.5. The luminosity functions show strong evolution. The brightest galaxies are 4 magnitudes brighter than their present day counterparts, and the faint galaxies are fewer in number. The double-peaked redshift distribution and the evolution of a luminosity function can be understood if larger galaxies form stars at z ~ 3, and if star formation is delayed in the dwarf galaxies until after z ~1.

A typical high-z galaxy with R=24.5 and z=3 has a far-UV luminosity L= 1.3 x$10^{41}$ $h_{70}^{-2}$erg s$^{-1}$ Å$^{-1}$ at 1500 Å ( with Hubble constant $h_0$=70 km/s). The UV luminosity at z~3 exceeds by a factor of 30 the most luminous local example of the Wolf-Rayet Galaxy NGC 1741, which contains $10^4$ O-type stars. The typical luminosity corresponds to a star formation rate of ~ 8 $h_{70}^{-2}$ $M_\odot$ yr$^{-1}$. This is a lower limit because the dust extinction and a lower age would raise this value. Pettini et.al. (6) estimates a star-formation intensity ~ 13 $M_\odot$ yr$^{-1}$ kpc$^{-2}$. The star-forming galaxies at high-redshift appear to be spatially more extended versions of the local starbursts. The same physical processes, which limit the maximum star formation intensity in nearby starbursts, also seem to be at play in galaxies at high redshifts.

The objects are typically of ultraviolet luminosity ~ $10^9$-$10^{10}$ times the solar luminosity, and have sizes ~ 1 kpc and comoving spatial densities which vary between 0.05 and 0.01 Mpc$^{-3}$ (at redshifts between 2.5 and 6). The UV luminosity and sizes are similar to those of the nearby starburst galaxies, while the co-moving spatial densities are comparable to that of present-day galaxies. The objects appear to represent star-formation in small concentrated regions rather than in galaxy-sized objects. At early epochs spanned by the objects, the dynamical timescale of the galaxy-sized objects is comparable to the age of the universe, which suggests that we may be witnessing the first star-formation associated with the initial "collapse of galaxies". These objects could be proto-galactic regions of star-formation associated with the progenitors of present day normal galaxies at very early epochs.

For the examples are 0000-D6, 0201-C6 and cB58, the redshift of the strong interstellar absorption lines is at z=2.96. The peak of the Lyα emission is at a relative velocity 800 km s$^{-1}$, while the red wing extends to ~ 1500 km s-1, and the blue wing is sharply absorbed. This P Cygni type profile can be understood as originating in an expanding envelope around the HII region. The unabsorbed Lyα photons are back-scattered from the receding part of the nebula. The systemic velocity of the star-forming region is ≈ 400 km s-1, as measured from the wavelength of weak photospheric lines from O stars. The relative velocities of the interstellar, stellar and nebular lines point to large scale outflows in the interstellar medium, as a consequence of the starburst activities in the galaxy. Generally less energetic outflows are seen in local starburst galaxies.

The large scale motion of this sort could be the main reason for the strengths of the interstellar lines in high-z galaxies. Such strong absorption lines are also often seen in damped Lyα systems, that are not strongly associated with sites of active star formation, and where they may reflect complex velocity fields of merging protogalactic clumps.

At z ~3 the nebular emission lines which dominate the optical spectra of the star-forming galaxies are redshifted into the infrared H and K bands. The line widths, which reflect the overall kinematics of the star-forming regions, can provide an indication of the masses involved; a detection of $H_\beta$ (or $H_\alpha$ at z < 2.5) would give a measure of the star-formation rate which can be compared with that from the UV continuum, and the ratios of the familiar nebular lines may provide the way of estimating the metallicity of the galaxies, given the complexity of the UV absorption line spectra.[6]

Powerful starbursts in the present universe emit almost all their light in the far-infrared rather than in UV. This may also be true at high redshifts. It is estimated [7] that an average vacuum-UV-selected galaxy at high redshift suffers 2 to 3 magnitudes of extinction. If the strong correlation between vacuum-UV colour and metallicity in local starbursts is applied to high-z galaxies, it would suggest a broad range of metallicity from substantially sub-solar to solar or higher values, and a median value of 0.3-0.5 solar values. This is significantly higher than the mean metallicity in the damped Lyα systems. If extinction corrections are applied, the high-z galaxies have large bolometric luminosities ( ~ $10^{11}$ to $10^{13}$ $L_\odot$). This bolometric surface brightness of the extinction-corrected high-z galaxies are very similar to the values seen in the local starbursts. The implied average surface-mass-density of the stars within the half-light radius is quite similar to the values in the present-day ellipticals.

One prediction based on the local starbursts is that the high-z galaxies should show a strong correlation between the strength of the UV absorption lines and β, the spectral slope in the vacuum-UV continuum (more metal rich local starbursts are both redder and stronger-lined).

*II.1. Spectra*
Star-forming galaxies with spectra broadly similar to those of present day starburst galaxies, have been detected up to redshifts ≈5 [8]. At this early epochs the intergalactic medium was fully ionized, galaxies were enriched in a wide variety of chemical elements (albeit in lower proportions than today), and structures on the scale of rich clusters were already evident. The lowest metallicity measured at z = 4-5 are ~ 1/300 solar. The ionizing photons associated with the production of even such an amount of heavy elements outnumber the baryons by 10 to 1. Thus, if the initial enrichment took place when the universe was mainly neutral, the associated Lyman continuum radiation was sufficient to reionize all the diffuse gas.

Recent observations of normal galaxies at redshifts $z \approx 3$ have demonstrated that the Lyman-limit spectral break and Ly$\alpha$-forest spectral decrement, which arise due to photoelectric absorption by neutral hydrogen along the line-of-sight, constitute the most prominent spectral signatures for the very distant galaxies. The Lyman break objects in the redshift range $3.0 \leq z \leq 3.5$ represent about 1.3% of all objects to 25.0 mag, and 2.0% of all objects in the magnitude range $23.5 \leq \mathcal{R} \leq 25.0$. The spectra of very-high-redshift galaxies are characterized by a complete absence of flux below the Lyman-limit, and strongly absorbed flux in the Ly$\alpha$ forest. These spectral signatures, imprinted by intervening material, must apply irrespective of the spectral properties of the galaxies.[9]

Despite the apparent lack of a large amount of dust, Ly$\alpha$ emission is always much weaker than the ionization-bound dust-free expectations. In most z>3 galaxies the weakness of the Ly$\alpha$ line is strikingly similar to nearby star-forming galaxies.

The Keck spectra of z>3 galaxies show similarity with local starburst galaxies. The dominant charateristics of the far-UV spectrum are: A flat continuum weak or absent Lyman $\alpha$ emission; prominent high-ionization stellar lines He II, C IV, Si IV and N V; and strong interstellar absorption lines due to low-ionization stages of C, O, Si and Al.

As to the strengths of the lines which are predominantly stellar in origin (CIV, SiIV and HeII), there exists a great variety. Moreover, these line-features are generally weaker than in the spectra of present-day starbursts. They are formed predominantly in the winds of massive stars, where both mass-loss rates and wind terminal velocities depend sensitively on metallicity.[10]

At redshift $z < 2.3$, the spectra exhibit no significant absorption by intervening material. They are similar to the redshifted spectra of present-day galaxies. At redshift $2.5 \leq z \leq 4$ the spectra are characterized by a strong flux in the F814W and F606W images (the spectral sensitivities of these images peak at 8140 Å and 6060 Å), detectable flux in F450W images (with the sensitivity peak at 4500 Å) and no detectable flux in F300W images (with the sensitivity peak at 3000 Å). At redshift $4 \leq z \leq 5.5$ the spectra are characterized by a strong flux in the F606W images and no detectable flux in the F450W and F300W images. At redshift z> 6 the spectra are characterized by a strong flux in the F814W images and no detectable flux in the F606W, F450W and F300W images. The spectra of the objects at z > 6 are consistent with redshifted Lyman-limit absorption of high-redshift galaxies.

**III. STARBURST GALAXIES IN THE LOCAL UNIVERSE**

In the nearby galaxies ( < 10 Mpc), the four most luminous starburst galaxies are M82, NGC 253, M 83 and NGC 4945. Together they form 25% of the total high-mass star formation in this region. The rate of high-mass star formation in the few-hundred-parsec-scale starbursts in M 82 exceeds the rate in the entire disk of the spiral galaxy M 101, which has a surface area approximately four orders of magnitude larger than the M 82. Thus both in terms of energy production and rate of high-mass star-formation, starburst galaxies are highly significant components of the present universe.

The interstellar absorption lines are significantly blue-shifted with respect to the systemic velocity of the galaxy in many local starbursts. The true galactic systemic velocity lies between the velocity of the UV interstellar absorption lines and the Lyα emission lines. This is due to the outflowing gas, which both produces the blue-shifted absorption lines and absorbs the blue side of the Lyα emission line. Thus a purely UV signature of outflowing gas is a blueshift of the interstellar absorption lines with respect to the Lyα emission line.

The vacuum-UV spectral regime ( 912 -3000 Å) is the energetically dominant spectral region for the hot stars that power starbursts. The vacuum-UV spectra of starbursts are characterized by strong absorption features. These absorption features can have different origins like stellar wind, stellar photospheres, and interstellar gas. The resonance lines, due to objects with low ionization potentials (OI, CII, SiII, FeII, AlII, etc.), are primarily interstellar in origin. In contrast, the resonance lines due to high ionization (NV, SiIV, CIV) can contain significant contribution from both stellar winds and interstellar gas. The most unambiguous detection of stellar photospheric lines is provided by excitation transitions, but these lines are usually weak.

The greater the fraction of the UV which is absorbed by dust and re-radiated in the far IR, the redder the vacuum-UV continuum. Much of the available data strongly suggests that much of the dust responsible for the vacuum UV extinction is apparently distributed around the starburst in the form of a moderately inhomogeneous foreground screen, or "sheet" surrounding the starbursts. The vacuum UV extinction correlates strongly with the bolometric luminosity of the starbursts. Only starbursts with $L_{bol}<$ few x $10^9$ $L_\odot$ have colours expected for an unreddened starburst and the vacuum UV luminosity that rival their far-IR luminosity. The starbursts which lie above $L_{bol}>$ few x$10^{10}$ $L_\odot$ have red continua and are dominated by far-IR emission.

Apart from the effect of the dust, the starburst's metallicity is the single most important parameter in determining the UV properties: At low metallicity a significant fraction of vacuum UV escapes starbursts, and the vacuum-UV colours are consistent with intrinsic(unreddened) colours expected for a starburst population. In contrast at high metallicities (> the solar), 90-99% of the energy emerges in the far-IR, and the vacuum UV colours are very red. This means that vacuum-UV radiation escaping from starbursts suffers an increasing amount of reddening and extinction as the dust-to-gas ratio in the starburst ISM increases with metallicity.

The properties of the vacuum-UV absorption lines are also strongly dependent on metallicity. The high-ionization CIV and SiIV, and low-ionization CII and OI and SiII resonance absorption lines are significantly stronger in starbursts with high metallicity. The metallicity dependence of the high-ionization lines can be understood from high likelihood of strong contributions from stellar "wind".

The metallicity dependent strengths of the UV absorption lines which are stellar photospheric, are generally rather weak or blended with strong absorption features. They include CIII, SiIII, SV, SiIII and FeIII. The strong interstellar lines are optically thick. Their strength is determined to first order by velocity dispersion in the starburst gas. The enormous strengths of the starburst interstellar lines require very large velocity dispersion in the absorbing gas (of a few hundred km s$^{-1}$). There is very little correlation between the strengths of the interstellar absorption lines and the rotation speed of the host-galaxy. The interstellar lines are often blueshifted by one-to-several hundred km s$^{-1}$ with respect to the systemic velocity of the galaxy, showing that the absorbing gas is flowing outward, feeding the "superwind".

The true galatic systemic velocity lies between the velocity of UV interstellar absorption lines and the Lyα emission line. This is due to the out-flowing gas that both produces the blue-shifted absorption lines and absorbs away the blue side of the Lyα emission line.

Nuclear out-flows have been observed in a large number of starburst galaxies, e.g., M82, NGC 253, NGC 1705, NGC 2782, NGC 5253, NGC 3628, NGC 3079, NGC 6240, Arp 220, etc.

*III.1. M82*

In M82 the star-formation rate is dramatically enhanced[11]. Many of the star-forming regions are dusty and difficult to trace optically, and with abundant warm clouds of dust, gas, and numerous HII regions, they are particularly suited for millimeter wavelength interferometry.

There exist compact radio features coincident with supernova remnants with ages of a few hundreds or thousands of years. A new remnant is thought to appear every twenty years. In surface brightness and diameter, these remnants are similar to, but younger and smaller than, the equivalent remnants in the LMC and our Galaxy. The cumulative number as a function of diameter appears to increase linearly to a diameter of at least 3 pc, and indicates a supernova rate of 0.05 per year if the shells are expanding at 5000 km/s.

Star-formation in M82 is moving outwards from the centre, and CO observations show a ring of molecular gas with a radius centred at 200 pc. This is the region of current star formation, and the radio SNRs appear to be situated on its inner rim. The central 180 pc around the inner core possesses a steeper radio spectral index than regions further out from the core. This steeper region lies within the ring of molecular gas which surrounds the nucleus.

Several of the SNRs show spectral flattening. In the extreme case, there is a strongly rising spectrum between 1.4 GHz and 5 GHz. The spectra for SNRs are steep between 5 GHz and 8.4 GHz. About 40% of the more compact SNRs show evidence of low-frequency turnovers. These sizes are too large to be explained by synchrotron self-absorption. Instead, free-free absorption in the gas surrounding such SNRs, can explain the low-frequency turnovers. Moreover, the random distribution of the remnants, exhibiting this behaviour, implies that the absorption is caused by local clumps of gas rather than one single cloud.

*III. 2. NGC 253*

The HI distribution in NGC 253 is asymmetric in the outer regions, probably due to strong conribution of C of the spiral arms. A bar associated with the disc is visible in the optical and near-infrared. HI absorption measurements reveal unusual motions of the gas in the nuclear region, which seems to indicate a fast rotating ring of cold gas as well as outflow of gas.

*III. 3. NGC 1705*

NGC 1705 is a nearby HI-rich dwarf galaxy, which is dominated by extended loops of H$\alpha$ emission. The continuum emission shows a major axis extent of ~ 40 arcsec, and a minor axis extent of ~ 20 arcsec, while the H$\alpha$ loops extend beyond 40 arcsec in the direction perpendicular to the minor axis. At least eight major loops and arcs have been catalogued from the H$\alpha$ images, and are reminiscent of HI supershell structures in our Galaxy. A super-star-cluster NGC 1705A is present in the centre ( with core diameter ~ 2 pc). The stellar velocity dispersion and mass ($10^5$ M$_\odot$) of this supercluster is consistent with young globular clusters. The presence of the extended HI loops and the central super-star cluster point to the existence of a superwind. [12] The kinetic energy supplied by starbursts' supernovae over the last $10^7$ yrs may have created a super-bubble with a diameter of several kpc, that expands with a speed of about 50 km/s.

Between 1300 and 1700 Å , the continuum can be represented by a power-law : F$_\lambda$ $\propto$ $\lambda^{-2.5\pm0.2}$. Spectra show complex multi-component absorption from all the common interstellar species. The strongest features are SiII, CIII, SII, FeII, AlII, CI, CI*, OI, SiIV, CIV. The CIII absorption feature is broad and centred at ~ 620 km/s. This CIII absorption features commonly seen in OB stars, is a photospheric line, and originates from the massive stars in the central super-cluster. The SiII is interstellar with three components: A strong feature centred at -20 km/s, a relatively weak feature centred at 260 km/s, and a strong feature centred at 540 km/s. The other lines also show similar velocity structure.

The absorption feature centred at -20 km/s can be attributed to the Milky Way disk and halo gas. The 260 km/s feature is closely associated with LMC and the Magellanic stream, which intercepts the NGC 1705 line-of-sight. The H$\alpha$ and OIII nebular emission lines are double-peaked with a separation of ~ 100 km/s over most of the face of the galaxy. The kinematics of this ionized gas is best represented by a homogeneously expanding ellipsoidal shell. The blue-shifted emission components of the kpc-scale expanding supershell is at 540 km/s. This component has strong SiII and AlII lines, but relatively weak FeII absorption. The Si/S ratio suggests low dust depletion in the super-shell. The low FeII column density for the 540 km/s cloud seems to be intrinsic.

Galactic winds from this dwarf occur at time scales less than the time required to produce a Type Ia supernova. Therefore a Type Ia Sne which produces nearly 2/3rd of the iron and iron peak elements (Fe, Cr and Zn), cannot contribute to the metallicity of this gas, produced by the first generation of the stars, which are expelled from the galaxy. Shells are enriched with Type II supernova products like Si and Al, and deficient in Fe and other iron-peak elements (which are mainly produced by Type Ia supernovae).

*III. 4. NGC 2782*

NGC 2782 is a nearby peculiar spiral galaxy with a pair of HI and optical tails, and three ripples within the disk. [13] An Hα map shows an arc of inner emission along the inner ripple at 4.2 kpc, and an unresolved bright region of star formation in the inner 1.7 kpc radius. This region harbours one of the most luminous circum-nuclear starbursts among nearby spirals with luminosity comparable to M82. The circum-nuclear region has optical emission spectra indicative of HII regions as well as an additional component of highly ionized gas. The high excitation gas lies in a blue-shifted outflowing emission nebula[14].

The starburst-driven outflow in NGC 2782 forms a well-defined collimated bubble which has an extent of ~ 1 kpc and a closed shell at its edge. In most outflows, Hα and nonthermal RC form complex sets of filaments and loops. The soft X-ray and RC emission generally extends further out than Hα. An outflow similar to NGC 2782 exists in the LINER galaxy NGC 3079, that has an Hα bubble of radius 1.1 kpc with a shell feature at its outer edge. ( However, NGC 3079 forms two highly fragmented halos which extend twice as far as the Hα bubble.)

NGC 2782 contains cold molecular gas in addition to hot and warm ionized gas. The distribution of the central region shows an elongated clumpy feature with a double-lobed (of CO peaks) bar-like shape of radius 1.3 kpc. Two CO spurs appear to originate from the centre of the CO bar where the starburst activity peaks. CO iso-velocity contours show kinks at the base of the spurs, indicating deviations from circular motion.[15] The spurs are elongated along the minor axis and inside the RC outflow bubbles, suggesting outflowing gas.

*III. 5. NGC 5253*

NGC 5253 is a dwarf galaxy in the Centaurus group[16]. Although the outer regions have properties similar to a dwarf elliptical, star formation is active in its central core. The core hosts about a dozen of blue stellar clusters, as well as diffusely distributed high-mass stars. The peak intensity of the star formation is concentrated in the Northern part of the core. The inhomogeneous structure and the intense star-formation in the central region suggests its classification closer to an amorphous galaxy or a blue compact dwarf, rather than an elliptical. The starbursts in the nuclei include Wolf-Rayet stars and a small number of supergiants. Radial inflow of peripheral HI gas has been suggested as a possible source of fuel for the current starburst.[17] An encounter with the spiral M83 may have triggered the star formation.

Over a scale of tens of arcsecs the Hα emission is roughly circularly symmetric with many loops and filaments extending radially off the central region. The most recent star formation is probably responsible for heating the gas which emits in the soft X-rays, as well as for the complex system of loops and filaments of ionized gas. The peak of the Hα emission is roughly located at the centre of the nuclear region, and coincides with the position of a stellar cluster. At NW of the Hα peak, a multiple shell structure with radii of about 20 pc each is visible. This is one of the positions where enhanced nitrogen enrichment has been measured. In the South of the complex, roughly circular structures of ionized gas of about 65 pc is present. Two shell nebulae of about 25 pc diameter are located about 125pc ENE and about 280 pc SSE of the core. The first shell ( the brightest of the two) contains two blue stellar objects. At the edges of the frame , at a distance of about 0.7-0.8 kpc from the core, two extended filamentary structures can be discerned ESE and W of the centre.

Intense photo- and shock ionization of the ionized gas and the star-formation seem to be active up to at least 400 pc from the core. Individual expanding bubbles and enhanced knots of emission exist hundreds of pc from the core.

## IV. EMISSION AND SPECTRA

*IV. 1. X-ray emission at (0.4-10.0 keV) from M82 and NGC 253*

The X-ray spectra of M82 and NGC 253 are very similar to the spectra of low-luminosity AGNs and LINERs which generally can be described by a Raymond-Smith plus power-law model. It is likely that some of the point sources have masses $\gg 10\ M_\odot$.

X-ray emission is concentrated both in the nucleus and extended along the minor axis to scales up to ~ 10 kpc. Much of the central emission is from point sources. The spectrum is complex with at least two components. The nuclear emission is due to an extended component, probably consisting of the emission of unresolved supernovae and SN-heated ISMs, and individual point sources which are very luminous ( and can be X-ray binaries or young supernovae). The plumes breaking out of the galactic disk give evidence for "superwind".

The soft extended halos are ( for kT< 0.5 keV) contributing negligibly in the 0.5-2.0 keV bandpass compared to the flux originating from the region of a 6' radius centred on the nuclear source. The double Raymond-Smith model gives a better fit to the M82 spectra, while the Raymond-Smith plus power-law model gives a better fit to the NGC 253 spectra.[18] The hard and soft components are absorbed well in excess of the galactic columns ( $1.6 \times 10^{20}$ cm$^{-2}$ for M82 and $4.3 \times 10^{20}$ cm$^{-2}$ for NGC 253). The soft-component temperatures in the two cases are similar ( at 0.6 and 0.8 keV), while the hard component appears to be harder in M82 ( with kT ~ 11 keV) than in NGC 253 (with kT ~ 7keV). The temperature in the disk and nuclear regions is ~ $10^7$ K, with temperature decreasing with radius out in the halos (beyond ~ 1 kpc). In NGC 253 the soft (<2 keV) component is apparently dominated by a disk and a nuclear thermal component with kT ~ 0.8 keV. In M82 the soft component appears to have approximately equal contributions from ~ 1.1 keV gas in the nuclear and disk regions, and ~ 0.4 keV gas in the halo. These are consistent with the starburst driven wind. The outflow velocity of the X-ray emitting gas in M82 may be as high as 1000-3000 km s$^{-1}$, although the outflow velocity of the C gas in M82 is estimated to be less than 500 km/s.

In these galaxies Fe abundance is suppressed, particularly relatively to S and Si. Since Fe is depleted in warm ISM clouds to a larger extent than S and Si, dust depletion may account for some of the Fe deficiency.

*IV.2. X-ray emission from LINERs and Starburst galaxies*

AGNs are compact (<1pc) variable sources, in which the X-ray emission has a non-thermal power-law form[19]. Seyfert galaxies are characterized by strong Fe K line emission produced by fluorescence in cold matter. On the other hand, nuclear starbursts are frequently extended over kpc scales, and are expected to have thermal spectra characterized by coronal X-ray emission lines.

X-ray spectra of LINERs (low-ionization emission-line regions) and starbursts are not very different to those of classical AGNs. The hard power-law slopes are similar to those of QSOs, as well as the inferred intrinsic slopes of the Seyferts. The quasars and low-luminosity spirals are devoid of large amount of matter in the nucleus, that is responsible for reprocessing of X-ray continua of the intermediate luminosity Seyferts. The soft extended thermal emission, which is common in many LINERs and starbursts, has been observed in classical AGNs, in which the hard power-law is heavily absorbed, allowing the soft component to be detected.

The X-ray emission fits well by a simple power-law plus absorber model, where multiple components are required. In general, spectra of the nuclear sources can be modelled with a power-law plus an optically thin thermal Raymond-Smith model. The power-law emission and excess absorption is generally associated with the compact nuclear source, whereas the soft emission component is extended. In some cases the soft components dominate the spectrum, in others a power-law dominates, while in others both components contribute significantly. In most cases, the best-fitting temperature is consistently in the range 0.6-0.8 keV. The elemental abundance in many cases are sub-solar, and the power-law photon index is typically ~ 1.7-2.0. In a few sources the hard power-law is absorbed by a column density greater than $10^{22}$ cm$^{-2}$.

Variability has been observed in both LINERs and starburst galaxies on timescales of weeks to years. This implies that a significant amount of the emission originates from a compact (<< 1pc) region. On the other hand, rapid variability on the timescales of a day or less is not observed ( since rapid variability is most common in objects with 2-10 keV luminosity in the range of ~ $10^{42}$-$10^{43}$ erg s$^{-1}$, but vanishes below ~ $10^{41}$ erg s$^{-1}$). Long term variability, by a factor of up to 1.7 corresponding to a 2-10 keV flux in the range 1.4-2.4 x $10^{-11}$ erg cm$^{-2}$ s$^{-1}$, and 2-10 keV luminosity in the range 4.2-7.4 x $10^{40}$ erg$^{-1}$, was found.

These results imply a connection of the soft component with the warm gas with kT ~ 0.6-0.8 keV, possibly from an SNR-heated ISM, and starburst-driven winds. In some cases a hard component may also be due to starburst activity, resulting from compact supernovae.

*IV. 3. Infrared iron emission-line:*

Iron is a refractory element heavily depleted on dust in the diffuse ISM. Fast shocks propagating in the diffuse ISM (> 100 km/s) as those produced by superwinds, can destroy dust grains through a sputtering process or grain-grain collisions, and replenish the ISM with gas-phase iron. The gaseous iron is then collisionally excited in the cooling post-shock gas, and produces the infrared emission.[20] Supernova remnants show an enhanced FeII/H ratio up to about 1000 times more than HII regions. In galaxies, the FeII emission appears to be positionally coincident with radio emission from SNRs and with regions of star formation.[21]

Most of the local starbursts have FeII/$H_2$ in the range 2-6 (The FeII is a measure of the "supernova" rate.) The nebular H-emission can be used to predict Type II supernova rates, because it gives a measure of the number of massive stars which are presently ionizing the gas. The total number of Type II supernovae produced by an instantaneous burst of star-formation, is at most a factor 2 lower than the total number of Type Ia supernovae, but is spread over a timescale about 30 times shorter. In a star-forming region, the SNII rate is about one order of magnitude larger than the SNIa rate, as observed for late type spirals.[22] In this case the FeII emission is produced mainly by shocks from SNII and is a measure of the star-formation activity in the recent past.

## V. STARBURSTS AND SUPERWINDS

*V.1. Causes of starbursts*

The low efficiency of the stars for energy production, coupled with severe energetic demands (ranging from $10^{59}$ ergs for a modest starburst like M 82 up to ~ $10^{61}$ ergs for an ultraluminous starburst like Arp 220), mean that the interstellar gas mass of at least $10^8$ to $10^{10}$ $M_\odot$ are needed to fuel a starburst (even assuming 100% efficiency for conversion of gas into stars). The surface mass density of the cold interstellar matter in starbursts is so high (typically $10^3$ $M_\odot$ pc$^{-2}$) that the growth time for gravitational instabilities that would lead to star-formation in the starbursts is extremely short (~ one million years). Moreover, the time scale for gas depletion in a starburst via star-formation and outflows are also short compared to the minimum time it would take gravity to move matter into the starbursts from the large-scale disk of the galaxy ($10^7$ to $10^8$ years versus $10^8$ to $10^9$ years). The gas, which fuels the starbursts must be assembled at least as fast as it is consumed. Since the gas has a mass comparable to the mass of the entire interstellar medium in a normal galaxy, powerful starbursts can only occur when some process allows a substantial fraction of the interstellar medium of a galaxy to flow inward by at least an order-of-magnitude in radius, at velocities that are comparable to the orbital velocities of the galaxy's disk.[23]

The collapse of a self-gravitating system implies a maximum infall rate of 25 ($v_{infall}$/50 km s$^{-1}$)$^3$ $M_\odot$ per year. This can be compared to typical estimated star-formation rates of 10 to 100 solar masses per year in starbursts and typical orbital velocities of roughly 150 km s$^{-1}$ in the host galaxy. Thus, fuelling of the starbursts requires a mechanism which can induce non-circular motions that are both large in amplitude and involve substantial fraction of the interstellar medium.

However, for an infall mechanism to work, the gas must lose angular momentum. The tidal stresses can perturb the orbits of the stars and gas in the disk. The dissipation of the kinetic energy when gas collides with other gas, allows the gas to be sufficiently displaced from the stars. Then, gravitational torques act between the stars and the gas, allowing significant amounts of angular momentum to transfer from the gas to the stars. Gas can thereby fall towards the centre, fuelling a starburst.

*V. 2. Superwinds*

Superwinds are galactic scale outflows. The temperature of the hot outflowing gas in a starburst is considerably cooler (a few to ten million degrees) than it would be expected for pure thermalized supernovae and stellar ejecta (at $10^8$ K). The gas temperatures in superwinds are observed to be independent of the galaxy rotation speed. Soft X-ray emission (hot gas) is a generic feature of the halos of the nearest starburst galaxies. The estimated thermal energy of this gas represents a significant fraction of the time-integrated mechanical energy supplied by the starbursts. Superwind may transport much of the mechanical energy supplied by high-mass stars into the IGM (Inter Galactic Medium). The gas flowing outward from the starburst probably feeds the superwind.

If superwinds carry substantial amounts of metal enriched gas out of the starbursts, one will see cumulative effects of these flows in the form of a metal-enriched IGM, and/or metal-enriched gaseous halos around the galaxies. Typical MgII absorption line systems at z<1, seen in the spectra of distant QSOs, arise in the metal enriched halos of the intervening galaxies (which are normal systems, and not starbursts). Nevertheless, it is plausible that the galactic halos/IGM may be polluted primarily by episodic eruptions associated with powerful starbursts.

The gas in the central region cools rapidly and begins to contract owing to the self-gravity of dark matter and gas. When the gas temperature becomes close to $10^4$K and stops decreasing, a quasi-isothermal contraction is established. A further increase in the gas density causes a burst of star-formation in the central region. Thereafter, as massive stars explode as SNe II, the surrounding gas acquires the thermal energy and the gas temperature rises up to about $10^6$ K. At the same time the gas is gradually polluted with synthesized metals. A shock wave propagates outward and the superwind drives outflow of gas from inside. This outflow collides with the infalling gas and high-density super-shell is formed. While the gas is continuously swept up by the super-shell, the gas density further increases due to an enhanced cooling rate in the already dense shell. Then the intense star-formation begins within the super-shell, and subsequent SN explosions accelerate the outward expansion of the shell. Star formation continues in the expanding shell (for about $10^8$ years) until the gas density in the shell becomes too low to form new stars. The oscillations of swelling and contraction of the system continues for several times $10^8$ years, and the system becomes settled in a quasi-steady state in $3\times10^9$ years. The resulting stellar system forms a loosely bound virialized system due to significant mass loss, and has a large velocity dispersion and a large core( and stars are formed before most of the gas is fully polluted by metals). Radial distribution of metal abundances in this system has a positive gradient which is in sharp contrast to the observed negative gradient for massive galaxies. This process of star formation turns out to reproduce the observed features of dwarf ellipticals. The heating by supernovae proves to be an ideal suppressing mechanism against the efficient formation of low-mass galaxies.[24]

## VI. SUPERNOVAE AND HYPERNOVAE

*VI. 1. "Failed Type I supernova/ hypernova"*

Supernovae may give rise to gamma-ray bursts when a shock breaks out through the stellar surface.[25] Strong adiabatic shocks accelerate small amount of matter to relativistic velocities, if the density of the ambient medium falls rapidly enough with distance. This highest velocity ejecta is followed by the bulk of mass expanding less rapidly, but carrying most of the kinetic energy. Woosley (32) proposed a "failed Type Ib supernova" model for GRBs. The massive, collapsed, rapidly rotating core may have 5-10 solar masses with a high-density disk spinning around it.

A model[26] where a spherical shock developed in a supernova explosion accelerates a small fraction of the outer stellar layers to highly relativistic velocities and gives rise to gamma-ray bursts, has been proposed. Unfortunately, radiative heat losses damp out the shock when it comes to the stellar surface, and there is no reason to expect a bulk relativistic flow and gamma-ray emission.

However, the idea that a small fraction of explosion ejecta reaches relativistic velocities may explain GRB characteristics. The bulk of the mass and the kinetic energy of the explosion may be stored in relatively slowly moving ejecta. When the leading shock slows down by sweeping up an ambient medium, the slower moving matter far behind the shock catches up and provides a powerful and long-lasting energy source for the afterglow.

The total kinetic energy of either Type I or Type II supernovae is ~ $10^{51}$ erg, and a small fraction of the kinetic energy of any explosion may be converted into gamma-rays, though much more powerful events like a "hypernova" will be necessary to explain the GRB phenomena.

Powerful streams of neutrinos and antineutrinos in "fireball models" in the context of a neutron star merger scenario, seem to fail to provide the mechanism for the generation of a fireball.[27] However, a magnetic field can provide sufficient coupling between the spinning core and torus and the stellar envelope. When a core of a massive star collapses, then the field may reach ~ $10^{15}$ Gauss. Right after its formation in a core collapse, the massive core may have up to 5 x $10^{54}$ erg stored in its rotation. If a large fraction of that energy is extracted rapidly by a strong magnetic field, then the kinetic energy of explosions may be ~ $10^{54}$ erg. If the total ejected mass is ~ 10 $M_\odot$, then all ejecta will be sub-relativistic. A small fraction of the ejected envelope mass may be accelerated to a very high Lorentz factor by the explosion, or by a strong shock expanding rapidly into the circumstellar envelope with a steep density gradient. These ultra-relativistic ejecta provide the conditions required for a "fireball" model.

In a "hypernova" scenario, the neutrino energy transport is replaced by magnetic transport. Any massive star has a strong wind throughout its evolution, and the hypernova ejecta will produce a relativistic shock in the ambient medium. At a sufficiently steep density gradient in the ambient medium, the shock may accelerate a small amount of mass to a very large Lorentz factor. The most dramatic difference of this model and the "fireball" model, involving neutron star merger, is in the long term behaviour of the afterglow. In a "hypernova" model the afterglow should last for a much longer time. Moreover, the "hypernova" scenario places GRBs in star forming regions, while any merging neutron star scenario places them far away from such regions. By the time the neutron stars merge, they have moved away from their place of original birth as a consequence of two consecutive supernova explosions.

*VI. 2. High-redshift supernovae*

High-redshift supernovae are the brightest class of supernovae. They rise to maximum light within a few weeks, and at high redshifts they fade below the largest telescopes' limits within a month or two. The redshift distribution depends on the telescopes' aperture and exposure times. Most of the high-redshift supernovae observed are in the range z=0.35-0.45. These SNe Type Ia are rare, rapid and random, and constitute a family of very similar events, though not all identical. This family can be described by a single parameter, especially representing the shape or width of the light-curve. This parameter is tightly correlated with the absolute magnitude at the maximum. The broad slow-light-curve supernovae appear somewhat brighter, while the narrow fast-light-curve supernovae are somewhat fainter. The width-brightness relation can be characterized by a magnitude correction $\Delta mag = 2.35 (1-s^{-1})$, where s is the stretch factor.

Observational evidence in classifying supernovae of type Ia is the following: A spectrum of the supernova is available, and it matches the spectrum of a low-redshift type Ia observed in the appropriate number of rest-frame days, past its light-curve maximum. This criteria will usually differentiate types Ib, Ic, and II from Type Ia. Near maximum light, supernovae Type Ia usually develop strong broad features, while the spectra of SN II are more featureless. Distinctive features such as a trough at 6150 Å (i.e., blueshifted Si II $\lambda 6355$) uniquely specify a SN Ia. The spectrum and morphology of the host galaxy can identify it as an elliptical or S0, since SN Ia are only found in these galaxy types (although SNe Ia are also found in late spirals at an even higher rate than in ellipticals locally). The light curve shape is also different from SN II.

*VI. 3. Supernovae Type Ia and cosmological time-dilation*

The Type Ia supernovae[28] are as a class highly homogeneous explosive events, which are apparently triggered under very similar physical conditions. Their light curves scatter by less than ~ 25% RMS in brightness, and less than 15% RMS in full-width-at-half-maximum. Their spectral signatures also follow a well-defined evolution in time.

A recent study[29] gives a single-parameter characterization of the Type Ia light-curve differences, particularly for the "non-normal" redder objects.[30] This parameter, a time-axis stretch factor s, stretches (or compresses) the template light curve. From this study based on nearby SNe, it has been shown that the stretch factor s extends over a range of 0.65 to 1.1. To observe the cosmological time-dilation experimentally, the expected dilation factor (1+z) is modified by the stretch factor s. The observable effect is then d=s(1+z). Since s is asymmetric around 1 for low values of "s", which occur for the most extreme 15% of non-normal Type Ia supernovae with red colours, the (1+z) effect can be essentially cancelled by s, giving an observed dilation of d ≈ 1. At redshifts between 0.1 and 0.5, examples of Type Ia supernovae with observed width greater than s (1+z)= 1.1 provide evidence for the time-dilation. Narrower supernovae neither help nor hurt in distinguishing models, since the two ranges both are consistent with supernovae with observed width smaller than 1.1. At redshifts higher than 0.5 all supernovae become useful for separating the models. The data for the majority of the distant SNe is consistent with cosmological time-dilation.

*VI. 4. GRBs and the star- formation in the starburst galaxies at high redshift*

A typical dim-GRB ( 1 ph $cm^{-2}$ $s^{-1}$ lasting 10s) at z=1 releases 3.5 x$10^{49}$ erg $sr^{-1}$ in gamma-rays. The engine behind this emission must provide 4.4 x $10^{52}$ $f_b$/ $\varepsilon_{-2}$ erg [31] (where $f_b$ is the fraction of the sky illuminated by gamma-rays, and $\varepsilon_{-2}$ is the efficiency of conversion of the initially available energy into gamma-rays). This is the natural energy scale of a supernova explosion, in which $10^{53}$ergs may be released. The bulk is carried away in neutrinos, and about 1% becomes the kinetic energy of the ejecta.

We propose that GRBs arise from a core-collapse of massive stars in star-forming regions. The details of the "failed-supernova/hypernova" model will be discussed in paper III. In this model the collapse of a massive star produces a dense torus of circumstellar material as well as a relativistic shock wave that generate shells around the collapsed core, which in turn causes high-energy gamma-ray emission and the afterglow.